\documentclass[prb,superscriptaddress,amsfonts,twocolumn,showpacs,floatfix]{revtex4}
\usepackage{amsmath}
\usepackage{amsfonts}
\usepackage{amssymb}
\usepackage{bm}
\usepackage{amsmath}
\usepackage{txfonts}
\usepackage{tabularx}
\usepackage[dvips]{graphicx,color}

\newcommand{\TN}{T_{\mathrm{N}}}
\newcommand{\pcr}{PdCrO$_2$}

\begin{document}
\title{
Critical behavior of the metallic triangular-lattice Heisenberg antiferromagnet PdCrO$_{2}$
}

\author{Hiroshi Takatsu}
\email{takatsu@scphys.kyoto-u.ac.jp}
\affiliation{Department of Physics, Graduate School of Science, Kyoto University, Kyoto 606-8502, Japan}

\author{Hideki Yoshizawa}
\affiliation{
Neutron Science Laboratory, Institute for Solid State Physics, 
The University of Tokyo, Tokai Ibaraki 319-1106, Japan
}

\author{Shingo Yonezawa}
\affiliation{Department of Physics, Graduate School of Science, Kyoto University, Kyoto 606-8502, Japan}

\author{Yoshiteru Maeno}
\affiliation{Department of Physics, Graduate School of Science, Kyoto University, Kyoto 606-8502, Japan}
\date{\today}

\begin{abstract}
We report physical properties of the conductive magnet PdCrO$_2$
consisting of a layered structure with a triangular lattice of Cr$^{3+}$ ions ($S=3/2$).
We confirmed an antiferromagnetic transition at $T_{\mathrm{N}}=37.5$~K
by means of specific heat, electrical resistivity, magnetic susceptibility,
and neutron scattering measurements.
The critical behavior in the specific heat persists in an unusually wide temperature range
above $T_{\mathrm{N}}$.
This fact implies that spin correlations develop even at much higher temperature than $T_{\mathrm{N}}$.
The observed sub-linear temperature dependence of the resistivity above $T_{\mathrm{N}}$
is also attributed to the short-range correlations among the frustrated spins.
While the critical exponent for the magnetization agrees reasonably with the prediction of the relevant model,
that for the specific heat evaluated in the wide temperature range differs substantially from the prediction.
\end{abstract}

\maketitle

\section{Introduction}
Geometrically frustrated spin systems have attracted much attention 
for the realization of novel ground states with unconventional order parameters.
Especially, antiferromagnetically interacting spins on two-dimensional (2D) triangular lattices (TL)
are actively investigated because the simple triangular arrangement of spins gives rise to 
various ground states
depending on the nature of the spins.\cite{Wannier1950, Mermin1966, Anderson1973, Teitel1983, Miyashita1984, Kawamura1984-1,Kanoda2006}
One remarkable example is a continuous spin system such as
XY or Heisenberg magnets.\cite{Teitel1983, Miyashita1984, Kawamura1984-1,Kanoda2006, Lee1984, Kawamura1984-2, Yosefin1985, Lee1986, Berge1986, Kawamura1998}
For such systems,
it is predicted that
the frustration is partially relaxed by forming a local 120$^{\circ}$ spin structure.
The short-range spin correlations 
produces a new degree of freedom called the vector chirality.
Interestingly, the vector chirality may exhibit a long range order
at a finite temperature
although the spins are \textit{not} long-range ordered.
However, it is difficult to detect the chirality order directly 
in real materials because 
many of them exhibit static long-range spin order due to additional interactions
(e.g.~inter-layer interactions)
above their predicted chirality-order temperature.
Nevertheless,
it is expected that 
the proximity to chirality order affects the critical
behavior of physical properties near a 
magnetic phase transition. \cite{Kawamura1998, Kawamura1987, Kawamura1992}
Indeed, e.\,g.\,,
the 2D triangular Heisenberg antiferromagnetic (THAF) compounds, V$X_2$ ($X=$~Cl, Br)
exhibit an unusual critical behavior with characteristic critical exponents. 
\cite{Kadowaki1987, Takeda1986, Wosnitza1994}
Therefore,
a detailed investigation of the critical behavior may reveal 
the interplay between spin and vector chirality.

Most of the 2D THAF compounds are insulators or semiconductors.
Therefore,
the magnetic properties of 2D metallic THAF compounds have not been sufficiently clarified.
In order to reveal the intrinsic interaction of the conduction electrons with frustrated spin moments,
it is desirable to investigate a clean metallic THAF compound without 
disorder introduced by chemical doping.

\begin{figure}
\begin{center}
 \includegraphics[width=7.5cm]{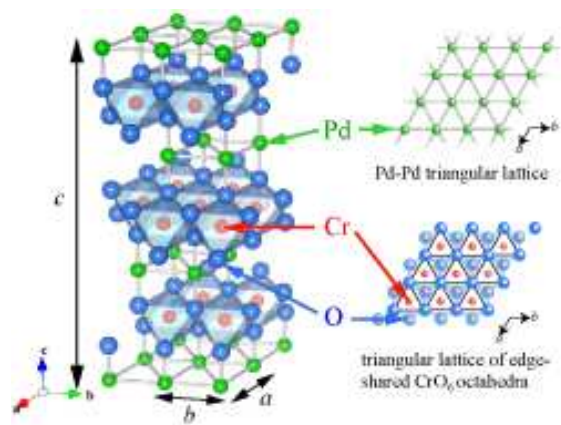}
\caption{
(Color online) 
Crystal structure of PdCrO$_{2}$.
This structure consists of alternating stacks of a triangular lattice of palladium
and a triangular lattice of edge-shared CrO$_{6}$ octahedra.
These drawings were produced using the software VESTA.\cite{Momma2008}
}
\label{fig.1}
\end{center}
\end{figure}

We studied the 2D THAF oxide 
PdCrO$_2$ (Cr$^{3+}$, $S=3/2$) and 
revealed that  
it exhibits metallic conductivity without chemical doping down to low temperatures.
Thus, we believe that
this oxide is one of the simplest and most suitable systems for the investigation
of the interplay between conduction electrons and frustrated magnetism,
as well as the involvement of the vector chirality.

PdCrO$_2$ crystallizes in the delafossite structure with the space group $R\bar{3}$m ($D^{5}_{3d}$).
This structure is closely related to the ordered rock-salt structure of
$A$CrO$_2$ ($A=$~Li, Na, K). \cite{Mekata1993}
However, they exhibit a different stacking sequence of the respective 
oxide and metal-ion layers in the unit cell.
In the delafossite structure, the noble-metal ions such as Pd$^{1+}$ are linearly
coordinated by two oxygen ions along the $c$ axis like a dumbbell,
whereas in the ordered rock-salt structure,
the alkali-metal ions $A^{1+}$ are zigzag connected to oxygen ions.
The lattice parameters of PdCrO$_2$ 
at 25$^{\circ}$C are $a=b=2.930$~\AA~and $c=18.087$~\AA.\cite{Shannon1971}
The temperature dependence of the magnetic susceptibility starts to deviate from a Curie-Weiss behavior 
below room temperature with a broad peak at about 60~K.\cite{Doumerc1986}
The Weiss temperature $\theta_{\mathrm{W}}$ and the effective moment $\mu_{\mathrm{eff}}$ were
reported to be $\theta_{\mathrm{W}} \simeq -500$~K and $\mu_{\mathrm{eff}} \simeq 4.1 \mu_{\mathrm{B}}$.
Moreover, a powder neutron scattering study down to 8~K revealed 
a magnetic transition around $T_{\mathrm{N}}=40$~K,
leading to a 120$^{\circ}$ spin structure. \cite{Mekata1995}
The frustration parameter $f\equiv|\theta_{\mathrm{W}}|/T_{\mathrm{N}}\simeq13$ 
indicates strong spin frustration.
To our knowledge the electrical resistivity and the specific heat of this compound 
have not been reported before our investigation.\cite{TakatsuHFM2008}
We note that the isostructural non-magnetic compound PdCoO$_2$
exhibits a metallic temperature dependence of the resistivity down to 20~mK, 
which is attributed to the Pd 4d$^9$ electrons.
It is also revealed that
the high-frequency phonons play an essential role in the temperature dependence
of the resistivity and specific heat.\cite{Takatsu2007}

This paper reports the critical behavior of metallic PdCrO$_2$ around its $T_{\mathrm{N}}$
investigated by detailed measurements of the specific heat.
We found that the resistivity of PdCrO$_2$
exhibits an unusual sub-linear temperature dependence
at temperatures above $T_{\mathrm{N}}$.
We also revealed the critical behavior in the specific heat 
persisting in the temperature range where
the $T$-sub-linear resistivity is found.
These results imply that the short-range correlation of frustrated spins 
gives rise to such characteristic behavior in PdCrO$_2$.

\section{Experimental}
\begin{figure}
\begin{center}
 \includegraphics[width=8.5cm]{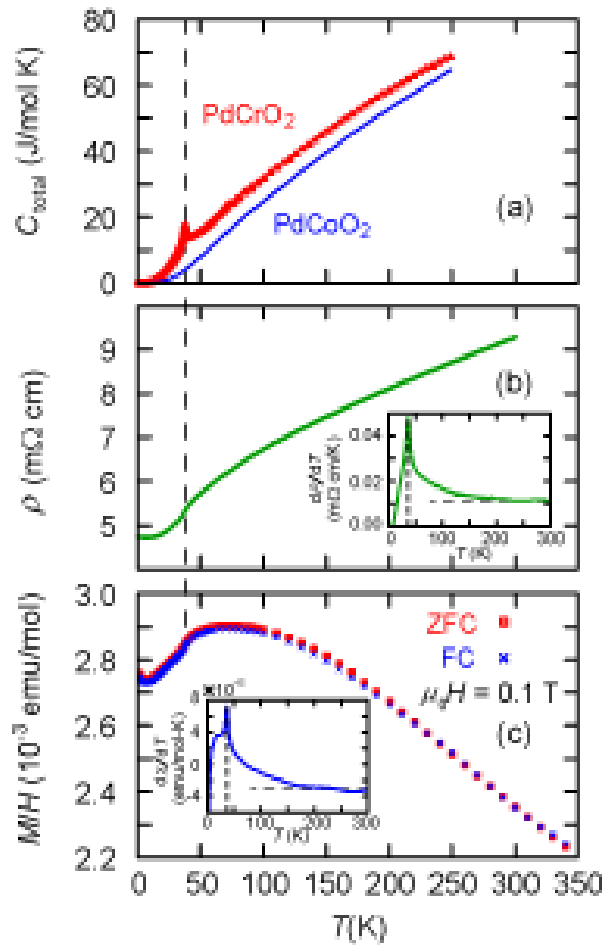}
\caption{
(Color online) 
Temperature dependence of (a) the specific heat, (b) electrical resistivity, and 
(c) magnetic susceptibility of PdCrO$_{2}$. 
The antiferromagnetic transition gives rise to anomalies at
$T_{\mathrm{N}}=37.5$~K in all these quantities.
The solid line in (a) represents the specific heat of 
the isostructural but non-magnetic compound, PdCoO$_2$ [Ref.~\onlinecite{Takatsu2007}].
The insets in (b) and (c) indicate the temperature derivatives of the resistivity
and susceptibility.
They exhibit a clear peak at $T_{\mathrm{N}}$.
Fig.~\ref{fig.2}(c) includes results of both zero-field cooling and field-cooling measurements of the magnetic
susceptibility at 0.1~T.
}
\label{fig.2}
\end{center}
\end{figure}
Powder samples of PdCrO$_2$ were obtained by the reaction, 
Pd + PdCl$_2$ + 2LiCrO$_2$ $\rightarrow$ 2PdCrO$_2$ + 2LiCl.\cite{Shannon1971,Doumerc1986}
We first prepared LiCrO$_2$ by reacting the stoichiometric mixture of 
LiCO$_{3}$ (99.99\%, Aldrich Chemical Co.) and Cr$_2$O$_{3}$ (99.99\%, Rare Metallic Co.\ Ltd.)
at 850$^\circ$C in an alumina crucible for 24 hours. 
We added Pd powder (99.99\%, Furuuchi Chem.\ Co.)
and PdCl$_2$ powder (99.999\%, Aldrich Chemical Co.) 
to the LiCrO$_2$ powder, and ground the mixture in a mortar for 30 -- 60 minutes.
It was then heated
to 790$^\circ$C and kept for 96 hours in evacuated quartz tubes.
The obtained samples were washed with aqua regia and distilled water to remove LiCl, 
the remaining Pd, and other by-products.
The samples were characterized by using powder X-ray diffraction (XRD) with CuK$_{\alpha 1}$ 
radiation and  energy dispersive X-ray analysis (EDX).
The XRD pattern indicated no impurity phase. 
The EDX spectra yielded a composition ratio of Pd and Cr of ${\rm Pd/Cr}=1.07$,
suggesting  stoichiometry within the experimental resolution.
However, 
the sample used in the powder neutron scattering experiment
contained an impurity phase as indicated by a neutron diffraction peak at
$2 \theta \simeq 53^{\circ}$, see Fig.~\ref{fig.Neutron_2theta}.
Since powder neutron diffraction requires 
a large amount of sample, we could not completely remove the impurity phase for that sample.

Specific heat was measured by applying a thermal relaxation method with a commercial calorimeter
(Quantum Design, PPMS) from 250 K to 0.35 K, using pressed pellets
sintered at 700~$^\circ$C for 12 -- 24 hours.
Around the transition temperature $T_{\mathrm{N}}$ we carefully took data points in $0.01$~K intervals. 
The temperature heating width $\varDelta\,T$
during each measurement was chosen to be $\varDelta\,T \leq 0.08$~K.
Note that a specific-heat measurement using the thermal relaxation method 
requires special attention to $\varDelta\,T$ in the immediate vicinity of a phase transition
because the obtained specific heat value is a certain average of $C_P(T)$ in the
temperature interval from $T$ to $T+\varDelta\,T$.

Electrical resistivity was measured with 
either AC or DC four-probe method from 300~K to 0.35~K.
In the temperature range from 2.0~K to 0.35~K, a $^3$He refrigerator (Oxford Instruments, Heliox)
was used.
Gold wires were attached with silver paste to rectangular samples cut out from pellets.

DC magnetization was measured with a commercial SQUID magnetometer 
(Quantum Design, MPMS) from 300~K to 1.8~K in magnetic fields $\mu_{0}H$
between 0.002~T and 7~T. 
The field dependence of the DC magnetization was also measured 
from $-7$~T to $7$~T at 4.2~K.

The powder neutron scattering measurements were performed in zero field on 
the triple axis spectrometer GPTAS-4G at JRR-3.
A pyrolytic graphite (002) reflection was used for the monochromator.
Higher-order neutrons were removed by a pyrolytic graphite filter.
The neutron wave length was fixed to $\lambda = 2.3475$~\AA,
and a collimation of 40'-- 80'-- 40'-- 80' was chosen.
The linear dimensions of the pelletized sample used was $23~{\mathrm{mm}} \times \phi~9~{\mathrm{mm}}$
and it was mounted in 
a $^3$He cryostat with a closed-cycle $^4$He refrigerator.
The present measurements extend to temperatures as low as 0.8~K,
compared with the previous report, 8~K. \cite{Mekata1995}
\section{Results}
\subsection{Specific heat}
The temperature dependence of the specific heat is
presented in Fig.~\ref{fig.2}(a).
The specific heat exhibits a clear peak at $T_{\mathrm{N}}=37.5$~K,
at which the temperature derivative of the resistivity (${\rm d}\rho/{\rm d}T$) and 
that of the susceptibility (${\rm d}\chi/{\rm d}T$)
also exhibit a clear peak.
Since previous \cite{Mekata1995} and the present neutron scattering measurements 
have also revealed magnetic Bragg peaks
at low temperatures, 
the results indicate that the static long-range spin-order occurs at $T_{\mathrm{N}}$.
It should be noted that the total specific heat ($C_P$) of PdCrO$_2$ is noticeably larger than
that of the isostructural non-magnetic PdCoO$_{2}$ 
even at 250~K, although it is often expected that $C_P$ of 
a magnetic compound approaches that of the non-magnetic counterpart at high temperatures.
This difference is mainly attributed to the difference of the Einstein modes in 
the phononic contribution ($C_{\mathrm{ph}}$).

In order to estimate the magnetic specific heat for PdCrO$_2$ ($C_{\mathrm{mag}}$),
we subtracted the phononic contribution ($C_{\mathrm{ph, Cr}}$) 
and the electronic contribution ($C_{\mathrm{ele}}$) 
from $C_{P}$ of PdCrO$_2$.
The phononic contribution $C_{\mathrm{ph, Cr}}$ was estimated by multiplying
$C_{\mathrm{ph}}$ of PdCoO$_{2}$
by a constant such that it approaches $C_P$ of PdCrO$_2$ at high temperatures;
we used 1.09 for this constant. 
This model empirically corrects for the number of the Debye modes
while effectively incorporating the reduction of the Einstein frequencies.
The electronic contribution $C_{\mathrm{ele}}$ ($=\gamma_{\mathrm{ele}}\,T$) was estimated by fitting the relation
\begin{alignat}{2}
 C_{P}-C_{\mathrm{ph, Cr}} &= C_{\mathrm{ele}}+C_{\mathrm{mag}} \notag \\
                          &= \gamma_{\mathrm{ele}}\,T + A_{\mathrm{mag}} T^{h}
\label{eq.cp}
\end{alignat}
to the data in the temperature interval between 0.35~K and 4.0~K.
In Eq.~({\ref{eq.cp}}), $\gamma_{\mathrm{ele}}$ denotes the electronic specific-heat coefficient, 
and $C_{\mathrm{mag}}=A_{\mathrm{mag}} T^{h}$ describes the magnon
contribution much below $T_{\mathrm{N}}$. \cite{Ramirez}
It is known that $h$ fulfills the realation
$h=d/\varepsilon$, where 
$d$ is the dimensionality of the magnon excitation and 
$\varepsilon$ is a parameter related to the type of the magnetic order
($\varepsilon = 1$ refers to AF order, 
$\varepsilon = 2$ to ferromagnetic order).\cite{Ramirez}
The coefficient $A_{\mathrm{mag}}$ is related to the spin velocity $v_{\mathrm{S}}$
in the long-range-ordered state. \cite{Ramirez1992, Nakatsuji2007}

The fitting yields
$\gamma_{\mathrm{ele}} = 1.4 \pm 0.2$~mJ/mol-K$^2$, $h = 2.0 \pm 0.1$,
and $A_{\mathrm{mag}}=1.2$~mJ/mol-K$^3$.
At 2~K, the magnitudes of the two terms (electronic and magnetic) are comparable, 
and the phonon contribution is only about 5\% of the total specific heat.
Thus, the electronic contribution can be evaluated with sufficient precision and accuracy.
The value of $\gamma_{\mathrm{ele}}$ is similar to the value found for non-magnetic PdCoO$_{2}$ 
($\gamma_{\mathrm{ele}}=1.28$~mJ/mol-K$^2$),\cite{Takatsu2007}
implying that the mass enhancement due to interactions between electrons and magnetic moments is not strong.
The observed $T^{2}$ dependence of $C_{\mathrm{mag}}$ 
is consistent with the theoretical prediction of the $T$-dependence 
of the specific heat due to 2D-magnon excitations in an AF ordered state,
i.\,e.\, $d=2$ and $\varepsilon=1$.
The spin velocity $v_{\mathrm{S}}$ is estimated to be 
$v_{\mathrm{S}}\simeq3200$~m/s
from the relation
$v_{\mathrm{S}}=[3.606\sqrt{3} R/2\pi A_{\mathrm{mag}}]^{\frac{1}{2}} (ak_{\mathrm{B}}/\hbar)$,
which is 
applicable to the magnetic order with 2D-magnon excitations in a TL magnet. \cite{Nakatsuji2007}

\begin{figure}
\begin{center}
 \includegraphics[width=7.9cm]{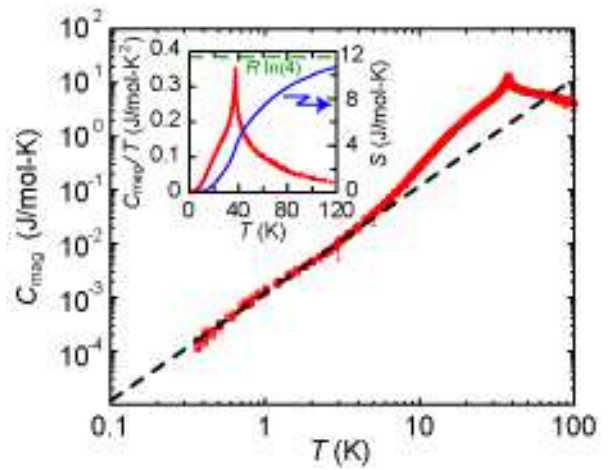}
\caption{
(Color online) 
Temperature dependence of the magnetic specific heat of PdCrO$_{2}$.
These data were obtained by subtracting the contribution of phonons
and conduction electrons. 
The phononic contribution was estimated from the specific heat of isostrucural but non-magnetic PdCoO$_{2}$.
The solid line is a fit to the data $\sim T^{2.0 \pm 0.1}$ indicating 
a realization of the 2D-magnon excitation in an antiferromagnetically spin-ordered state.
In the inset the temperature dependence of the magnetic specific heat divided by temperature 
$C_{\mathrm{mag}}/T$ is shown.
A sharp peak clearly indicates the transition at $T_{\mathrm{N}}=37.5$~K.
}
\label{fig.Cp_mag_1}
\end{center}
\end{figure}

The estimated $C_{\mathrm{mag}}$ (~=~$C_{P}-\gamma_{\mathrm{ele}}\,T-C_{\mathrm{ph}}$)
is displayed in Fig.~\ref{fig.Cp_mag_1}
on a logarithmic scale (main panel), which indicates a clear $T^{2}$ dependence below 4~K.
Interestingly, a small hump structure is observed at about 20~K below the AF transition
(inset of Fig.~\ref{fig.Cp_mag_1}).
A similar hump is also observed in CuCrO$_{2}$ and
some frustrated spinel magnets. \cite{Okuda2008, Tristan2005}
A recent theory 
points out that such a hump in $C_{\mathrm{mag}}$ below the magnetic transition
originates from quantum fluctuations of frustrated spins. \cite{Bernier2008-condmat}

The magnetic entropy at $\TN$ evaluated from the integration of $C_{\mathrm{mag}}(T)/T$
from 0.35~K is $3.9\pm0.1$~J/mol-K.
This value is remarkably small because it is only one third of 
the expected entropy for a system with $S=3/2$ localized spins: $R~\ln(\,2\,S+1\,)=11.53$~J/mol-K.
We have confirmed that the value of the magnetic entropy at $\TN$ is not affected by
the choice of a model for the estimation of $C_{\mathrm{ph}}$.\cite{TakatsuHFM2008}
The inset of Fig.~{\ref{fig.Cp_mag_1}} indicates that the entropy release persists 
to much higher temperatures.
This fact implies that strong short-range spin correlations persist to temperatures much above $T_{\mathrm{N}}$.

\begin{figure}
\begin{center}
 \includegraphics[width=7.9cm]{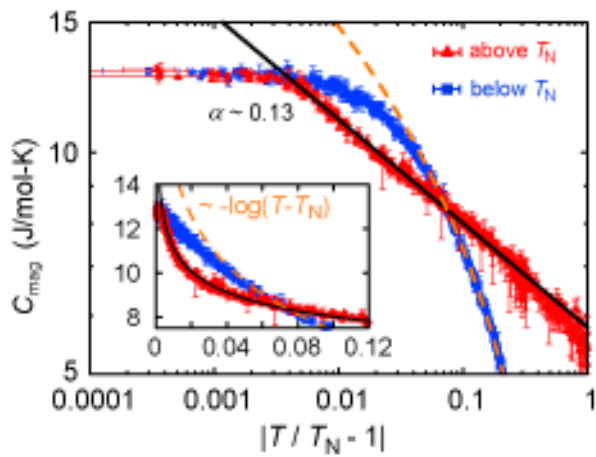}
\caption{
(Color online) 
Critical behavior of the magnetic specific heat.
The temperature range in which the critical behavior is observed extends up to 
around $T/T_{\mathrm{N}}-1=0.6$ (around 60~K).
The critical exponents $\alpha^{\pm}$ are estimated as 
$\alpha^{+}=0.13\pm0.02$ above $T_{\mathrm{N}}$ and as
$\alpha^{-}=0$ below $T_{\mathrm{N}}$.
}
\label{fig.Cp_mag_3}
\end{center}
\end{figure}

Figure~{\ref{fig.Cp_mag_3}} shows the behavior of $C_{\mathrm{mag}}$ 
in the critical region above and below $T_{\mathrm{N}}$,
displayed against the reduced temperature $t\equiv|T/T_{\mathrm{N}}-1|$.
The behavior in the critical region above $T_{\mathrm{N}}$ surprisingly
extends up to about $t\simeq0.6$; for ordinary magnets,
such behavior is observed only up to $t\simeq0.1$. \cite{Kornblit1973, Hoeven1968}
This behavior implies again that the magnetic correlations start to develop from temperatures
much higher than $T_{\mathrm{N}}$.
In order to estimate the critical exponents above and below $T_{\mathrm{N}}$,
we performed the fitting to the data with the relation,
\begin{alignat}{1}
C_{\mathrm{mag}} = A^{\pm}\, |T/T_{\mathrm{N}}-1|^{-\alpha^{\pm}},
\label{eq.C_critical}
\end{alignat}
where $\alpha^{+}$ and $\alpha^{-}$ denote exponents above and below $T_{\mathrm{N}}$, 
and $A^{\pm}$ represents respective coefficients.
The resulting exponent is 
$\alpha^{+} = 0.13 \pm 0.02$ ($0.005<t\le0.6$) above $T_{\mathrm{N}}$.
In the vicinity of $\TN$,
we reproducibly observed saturation of the divergence in several samples investigated in this study.
Such saturation is often observed in other magnets as well.\cite{H.E.Stanley}
Anticipating another value of $\alpha$ in the conventional critical region, 
we fitted Eq.(\ref{eq.C_critical}) to the data between 0.005 and 0.1.
The obtained value $\alpha^{+} = 0.14 \pm 0.02$ is essentially the same 
as that obtained in the wide temperature range.
This indicates, as it is also evident from Fig.~{\ref{fig.Cp_mag_3}}, 
that the critical behavior near $\TN$ persists to high temperatures.
For temperatures below $\TN$,
the critical behavior is weaker than $\log(t)$.
It is known that the three cases, a logarithmic divergence, a cusp and a jump at $\TN$,
are represented by  the critical exponent $\alpha = 0$. 
Thus we assign $\alpha^{-} = 0$
for the cusp observed in PdCrO$_2$.
We will further discuss the critical exponents $\alpha^{\pm}$ in Sec.~\ref{discuss}.

\subsection{Electrical resistivity}
In Fig.~{\ref{fig.2}}(b) the electrical resistivity of PdCrO$_2$ between 300~K and 0.35~K is shown.
The anomaly due to the AF transition is observed at temperatues centered around $37.5\pm0.5$~K
at which ${\rm d}\rho/{\rm d}T$ exhibits a clear sharp peak 
in agreement with the specific-heat data. 
The resistivity exhibits a metallic temperature dependence down to 0.35~K
but a very weak upturn of about 0.2\% below 7~K.
This weak upturn is attributed to a weak localization of conduction electrons
due to  grain boundaries in the pelletized samples,
because no anomaly was found at that temperature in other quantities.

By fitting the formula $\rho(T)=A+BT^{n}$
to the data below and above $T_{\mathrm{N}}$,
we obtained $n =2.9 \pm 0.1$ ($10~\mathrm{K}<T<35$~K) and 
$n =0.34 \pm 0.05$ ($38~\mathrm{K}<T<150$~K).
Below the transition temperature, 
the behavior in \pcr\ is consistent with that of
ordinary magnetic metals with localized moments,
which are known to exhibit such super-linear temperature 
dependence with $n>1$.\cite{Ziman1,T.Moriya,Kasuya1959}
However,
above $T_{\mathrm{N}}$, 
the observed sub-linear temperature dependence with $n<1$ is clear contrast
to the resistivity of ordinary magnetic metals. \cite{Kasuya1956, Mori1980}
We will discuss the origin of this unusual $T$-sub-linear resistivity observed in PdCrO$_2$
in Sec. \ref{discuss}.

We note here that there exists sample dependence in the absolute values of the resistivity, 
probably due to the fragility of the samples.
However, 
the qualitative behavior of the resistivity is consistent for all the samples investigated.

\subsection{Magnetic susceptibility}
The results of the DC magnetic susceptibility ($M/H=\chi$) 
for zero-field cooling (ZFC) and field cooling (FC) runs at 0.1~T
from 350~K to 1.8~K
are shown in Fig.~\ref{fig.2}(c).
The susceptibility $\chi$ continuously increases with decreasing temperature. 
Below about 200 K it starts to deviate from ordinary Curie Weiss behavior.
It exhibits a broad peak around 60~K as previously reported. \cite{Doumerc1986}
Such behavior is also reported for other 2D-THAF compounds, e.\,g.,
VCl$_{2}$ [Ref.~\onlinecite{Hirakawa1983}] and $A$CrO$_{2}$ ($A$=Li, Na),\cite{Olariu2006, Alexander2007}
indicating a development of 2D short-range spin correlations with decreasing temperature.
At $T_{\mathrm{N}}$, $\chi$ exhibits a continuous anomaly
and ${\rm d}\chi/{\rm d}T$ exhibits a clear peak.
Moreover, ${\rm d}\chi/{\rm d}T$ exhibits a shoulder structure around 20~K (the inset of Fig~{\ref{fig.2}}(c)),
at which $C_{\mathrm{mag}}$ also exhibits the hump.

Fitting $\chi(T)=\chi_{0}+C/(T-\theta_{\mathrm{W}})$ to the corrected data
in the temperature interval 250~K to 350~K,
we obtained 
the effective moment to be $\mu_{\mathrm{eff}}= 3.8-4.1 \mu_{\mathrm{B}}$ 
and the Weiss temperature to be $\theta_{\mathrm{W}}\simeq -500$~K, which 
are consistent with the earlier report.\cite{Doumerc1986}

\subsection{Neutron scattering}
\begin{figure}
\begin{center}
 \includegraphics[width=8.5cm]{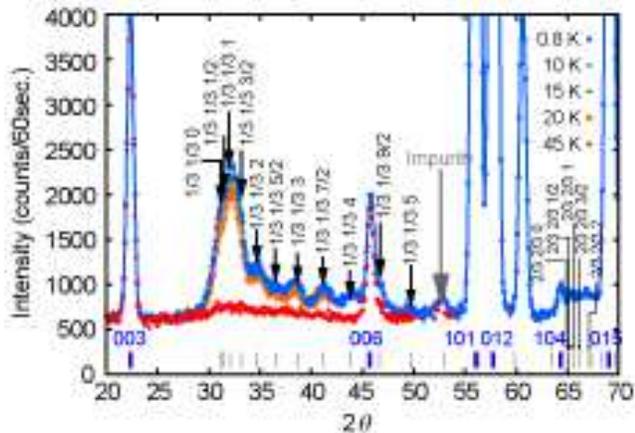}
\caption{
(Color online) 
Powder neutron diffraction patterns obtained at 0.8~K, 10~K, 15~K, 20~K, and 45~K
in zero field with the counting time of 60~sec. for each data point.
The peak positions are indexed.
The observed magnetic Bragg peaks are clearly broader than the instrument resolution.
The diffuse scattering feature centered around $\theta\sim32^{\circ}$ remains visible above $T_{N}$.
An impurity peak was detected at $2\theta=53^{\circ}$, see text.
}
\label{fig.Neutron_2theta}
\end{center}
\end{figure}
Powder neutron diffraction patterns at several temperatures in zero field are shown 
in Fig.~{\ref{fig.Neutron_2theta}}.
Below $T_{\mathrm{N}}$, we observed magnetic Bragg peaks.
The observed magnetic Bragg peaks are rather broad,
implying short coherence of the long-range order.
The peak positions can be labeled as ($\frac{1}{3}, \frac{1}{3}, l$) 
and ($\frac{2}{3}, \frac{2}{3}, l$) with $l=0, \frac{1}{2}, 1, \frac{3}{2}, 2 \cdots$,
which are consistent with the peaks for the 120$^{\circ}$ spin structure 
as previously reported by Mekata \textit{et al}. \cite{Mekata1995}
The intensity of each peak is also consistent with their report.
From the weaker relative intensity of the ($\frac{1}{3}, \frac{1}{3}, 0$) peak
compared with that of the ($\frac{1}{3}, \frac{1}{3}, 1$) peak,
they deduced that the spins lie in the $ab$ plane.
Interestingly, 
magnetic Bragg peaks with half-integer $l$ are observed 
in addition to the integer $l$ peaks. 
The spectrum observed in the delafossite PdCrO$_2$ 
resembles that of LiCrO$_{2}$ in the ordered rock-salt structure [Ref.~\onlinecite{Kadowaki1995}],
but in contrast with that of CuCrO$_{2}$ in the delafossite structure [Ref.~\onlinecite{Kadowaki1990}].
The observation of the half integer peaks suggests a LiCrO$_{2}$-type 
magnetic structure, which is believed to have two independent propagation vectors. \cite{Kadowaki1995}
Alternatively, the spectra are equally well interpreted in terms of two
kinds of magnetic domains with $l=$~integers and $l=$~half-intergers. 
In this case, the volume fractions of the domains contained 
in our sample are nearly the same.

As seen in Fig.~{\ref{fig.Neutron_2theta}},
the diffuse scattering feature between $2 \theta = 30^{\circ}$ and $44^{\circ}$
is still visible at 45~K (above $T_{\mathrm{N}}=37.5$~K).
This result indicates that 
strong short-range spin correlation is present at temperatures substantially higher than $T_{\mathrm{N}}$.
Note that such a diffuse component above $T_{\mathrm{N}}$ was also observed in other triangular lattice 
chromium compounds CuCrO$_2$, LiCrO$_2$, and NaCrO$_2$ \cite{Kadowaki1990, Soubeyroux1979}.

\begin{figure}
\begin{center}
 \includegraphics[width=7.9cm]{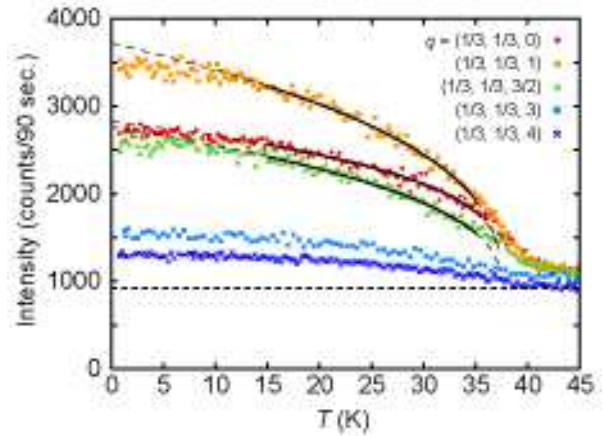}
\caption{
(Color online) 
Temperature dependence of the peak intensity of the powder neutron diffraction 
at several $q$ positions. The counting time is 90~sec. except for $\bm{q}=(\frac{1}{3}, \frac{1}{3}, 4)$,
for which it is 180~sec. 
The dotted line indicates the background, $I_{\bm{q}}^{\mathrm{B}}=920$~counts/90~sec.
The curves represent the results of the fitting between 15~K and 35~K with Eq.~({\ref{eq.nutron1}})
for the peaks at $\bm{q}=(\frac{1}{3}, \frac{1}{3}, 0)$, $(\frac{1}{3}, \frac{1}{3}, 1)$, and 
$(\frac{1}{3}, \frac{1}{3}, \frac{3}{2})$.
The resulting critical exponent lies within the value $\beta=0.18\pm0.03$.
}
\label{fig.Neutron_temp_dependence}
\end{center}
\end{figure}

The temperature dependence of the peak intensity at 
several $\bm{q}$ positions is shown in Fig~{\ref{fig.Neutron_temp_dependence}}.
Since the peak intensity at ($\frac{1}{3}, \frac{1}{3}, 4$) is very week,
the counting time was doubled to 180 seconds compared to the other measurements.
All the intensities exhibit a gradual increase with decreasing temperature 
and no anomaly around 7~K or 20~K.
These results indicate that there are no significant changes 
in the magnetic structure at low temperatures.

We estimated the critical exponent $\beta$ from the temperature dependence of the peak intensity, 
$I_{\bm{q}}(T)$,
at $\bm{q}=(\frac{1}{3}, \frac{1}{3}, 0)$, $(\frac{1}{3}, \frac{1}{3}, 1)$ and 
$(\frac{1}{3}, \frac{1}{3}, \frac{3}{2})$
using the relation,
\begin{equation}
M_{\bm{q}}(T)^{2} \propto I_{\bm{q}}(T) - I_{\bm{q}}^{\mathrm{B}} 
              = A(\bm{q})\,\Bigl|1-\frac{T}{T_{\mathrm{N}}} \Bigr|^{2\beta},
\label{eq.nutron1}
\end{equation}
where $M_{\bm{q}}(T)$ denotes the magnetization with wave number $q$,
$I_{\bm{q}}^{\mathrm{B}}$ represents the background intensity, and
$A(\bm{q})$ is the coefficient of the peak intensity at each $\bm{q}$.
We employed here the intensity at $2\theta \simeq 25^{\circ}$ (Fig.~{\ref{fig.Neutron_2theta}})
as the background, since the intensity at this position does not depend on temperature.
To estimate $I_{\bm{q}}^{\mathrm{B}}$ for the data in Fig.~{\ref{fig.Neutron_temp_dependence}},
we used the value multiplied by 1.5 to compensate for the different counting time 
between the $2\theta$ scans (counts/60~sec.) and the temperature scans (counts/90~sec.).
We used the fitting range from 15~K to 35~K in order to avoid 
the rounding region which is affected by the short-range spin fluctuation.
The fitting yields $0.18\pm0.03$.
This value is somewhat different from the previously reported value of $\beta=0.29$,\cite{Mekata1995}
probably due to a different choice of $\TN$;
in the previous report, the authors used $\TN=40$~K,
which was estimated from their neutron data itself,
whereas we used $\TN=37.5$~K at which a clear anomaly is observed in other physical quantities.
We believe that present choice of $\TN$ is more appropriate.

\section{Discussion}
\label{discuss}
In this section,
we will first discuss the unusual critical behavior observed in the specific heat 
and neutron scattering measurements for PdCrO$_2$.
Next we focus on the electrical resistivity exhibiting the characteristic sub-linear temperature 
dependence at temperatures above $T_{\mathrm{N}}$.
\subsection{Critical behavior}
Table \ref{table_exponents} summarizes the obtained critical exponents for PdCrO$_2$ 
and predicted values for various spin models.
The exponent $\alpha$ of PdCrO$_2$ above $T_{\mathrm{N}}$ ($t\le0.6$)
is not consistent with the predicted value for the layered THAF, SO(3). \cite{Kawamura1998}
Although the observed value is similar to that for the three-dimensional (3D) Ising-spin system,
this model is not appropriate for PdCrO$_2$ with Heisenberg spins in the quasi-2D crystal structure.
It should also be noted that the observed values of $\alpha$ above and below $T_{\mathrm{N}}$
are clearly different.
This is not consistent with the static scaling theories of the critical phenomena,
which predict that both $\alpha$ above and below 
the transition temperature should be identical. \cite{D.J.Amit, H.E.Stanley}
This discrepancy is attributable to the extended critical region to temperatures above $\TN$
due to spin frustration.
The observed value of the exponent $\beta$, which is closely related to the staggered magnetization below $\TN$,
agrees with those observed in V$X_2$ ($X=$~Cl, Br). \cite{Kadowaki1987,Takeda1986}
The main origin of the difference in $\alpha$ between PdCrO$_2$ and V$X_2$ is probably due to 
the much narrower temperature range used for the fitting for V$X_2$.
In principle, 
one should distinguish the short range fluctuation observed in the wide temperature range 
above $\TN$ from the critical fluctuation near $\TN$.
However, our experimental results suggest that they are not separable in PdCrO$_2$.

\begin{table}[t]
\begin{center}
 \caption{Critical exponents $\alpha$ and $\beta$ of PdCrO$_{2}$ and 
 the predicted values for various models.\cite{Kawamura1998, Guillou1977, H.E.Stanley}
 The symbol SO(3) denotes the order-parameter space applicable to the case of the layered THAF.
 The exponents for V$X_2$ ($X=$~Cl, Br) are also presented.
 The symbol $*$ represents a cusp at $\TN$.}
 \begin{tabular*}{0.46\textwidth}{@{\extracolsep{\fill}}lcc}
  \hline
                                   & $\alpha$       & $\beta$       \rule{0mm}{4mm} \\ \hline
   \, PdCrO$_2$ (above $\TN$)      & $0.13\pm0.02$  &   --          \rule{0mm}{4mm} \\
   \, \qquad\quad\, (below $\TN$)  & $0^{*}$            & $0.18\pm0.03$ \rule{0mm}{4mm} \\
   \, Ising (2D, square lattice)   & $0$~$(\log)$   & $0.125$       \rule{0mm}{4mm} \\
   \, Ising (3D)                       & $0.11$         & $0.325$       \rule{0mm}{4mm} \\
   \, XY (3D)                          & $-0.01$        & $0.345$       \rule{0mm}{4mm} \\
   \, Heisenberg (3D)                  & $-0.12$        & $0.365$       \rule{0mm}{4mm} \\
   \, SO(3)                            & $0.24$         & $0.30$        \rule{0mm}{4mm} \\
   \, VCl$_2$ (Ref.\onlinecite{Kadowaki1987}) & --             & $0.20\pm0.02$ \rule{0mm}{4mm} \\
   \, VBr$_2$ (Ref.\onlinecite{Takeda1986}) (above $\TN$)       & 0.59           & --            \rule{0mm}{4mm} \\
   \, \qquad\qquad\quad\,\,\,\, (below $\TN$) & 0.28           & $\sim0.20$    \rule{0mm}{4mm} \\
  \hline
 \end{tabular*}
 \label{table_exponents}
\end{center}
\end{table}

The critical behavior extending in
wide temperature range above the magnetic transition temperature 
is observed not only in PdCrO$_2$ but 
also in other 2D-TL magnetic systems. \cite{Alexander2007, Ajiro1988, Takeya2008}
Interestingly, in such frustrated systems, diffuse neutron scattering is also visible
in a similar temperature range. \cite{Kadowaki1990, Soubeyroux1979}
This fact manifests that
the critical state extending in a wide temperature range is one of the key features of the 2D-TL magnets.

\subsection{Metallic conductivity of PdCrO$_2$}
Here we discuss the temperature dependence of the 
electrical resistivity of PdCrO$_2$,
especially focusing on the $T$-sub-linear resistivity above $T_{\mathrm{N}}$.
For ordinary magnetic metals with non-frustrated localized spins,
the contribution to the  resistivity due to magnetic scattering
between conduction electrons and the localized spins above their magnetic transiton temperature 
is expected to be $T$-independent. \cite{Kasuya1956, Mori1980}
Therefore,
the temperature dependence of 
the total resistivity above the magnetic transition temperature 
is $T$-linear mainly due to electron-phonon scattering. \cite{Mori1980}
Compared with such a case, the $T$-sub-linear resistivity 
observed in PdCrO$_2$ above $T_{\mathrm{N}}$ 
is unusual, and it indicates gradual reduction of the magnetic scattering
from temperatures much above $\TN$.
In the temperature range in which the $T$-sub-linear resistivity is found,
the critical divergence is also observed in the magnetic specific heat
and deviation from the Curie-Weiss behavior in the susceptibility.
If one compares the insets of Fig.~\ref{fig.2}(b) and (c),
${\rm d}\rho/{\rm d}T$ and ${\rm d}\chi/{\rm d}T$
start to change at almost the same temperature.
Moreover,
the diffuse component in the neutron diffraction is observed 
above $T_{\mathrm{N}}$, which suggests the development of short-range spin correlations.
Thus, 
the unusual $T$-sub-linear resistivity of PdCrO$_2$ observed above $T_{\mathrm{N}}$
is attributable to the gradual development of short-range spin correlations,
which may reduce the spin randomness and weaken the magnetic scattering of the conduction electrons.

The presence of conduction electrons may further lead to long-range interactions between spins, 
such as the Ruderman-Kittel-Kasuya-Yosida (RKKY) interaction. 
Such interactions may in turn compete with the short-range exchange interaction and affect the spin frustration.
However, in PdCrO$_2$, the conduction electons do not seem to strongly affect the spin frustration, 
as evidenced by the 120$^{\circ}$ spin structure below $\TN$ and the value of beta comparable to those of 
the corresponding insulators.

\section{Conclusion}
In conclusion,
we have investigated the 2D-HTAF compound PdCrO$_2$
by means of specific heat, electrical resistivity, 
magnetic susceptibility, and neutron scattering measurements.
We found that PdCrO$_2$ exhibits metallic conductivity down to 0.35~K
without chemical doping.
We also confirmed the antiferromagnetic transition at $T_{\mathrm{N}}=37.5$~K
in the specific heat, which also affects the electrical resistivity.
Powder neutron scattering measurements revealed 
that this compound forms a $120^{\circ}$ spin structure 
with magnetic Bragg peaks of ($\frac{1}{3}, \frac{1}{3}, l$)
and ($\frac{2}{3}, \frac{2}{3}, l$) 
with $l=0, \frac{1}{2}, 1, \frac{3}{2}, 2 \cdots$ below $T_{\rm N}$, 
similar to LiCrO$_2$.
Contributions due to diffuse scattering are strongly present even above $T_{\mathrm{N}}$,
which implies that the short-range spin correlations start to develop at temperatures much higher
than $T_{\mathrm{N}}$. 
Moreover, 
the magnetic Bragg peaks are broad even at temperatures much below $T_{\mathrm{N}}$.
This fact implies that coherence length of the ordered moments remain rather short.

In the magnetic specific heat 
we observed a critical behavior that extends in an unusually wide 
temperature range above $T_{\mathrm{N}}$.
Moreover,
the critical exponents do not match with the exponents of standard models,
and these are also strongly asymmetric above and below $T_{\mathrm{N}}$.
Such an extended and asymmetric critical region
is attributed to the behavior characteristic of 
the frustrated Heisenberg spins on a triangular lattice.

In the electrical resistivity,
we observed a sub-linear temperature dependence above $T_{\mathrm{N}}$,
which is quite different from the resistivity of ordinary magnetic metals with non-frustrated localized spins.
This behavior is also characteristic of the conductive magnet with frustrated spins.

From the present study,
it is still unclear whether or not the metallic conductivity affects the unusual critical phenomena.
However, we found that the frustrated spins do affect the metallic conductivity of PdCrO$_2$.
Single crystal experiments in the future may allow us to extract 
the subtle effects of conduction on the geometrically frustrated spins.

\section*{Acknowledgment}
We would like to thank K. Ishida, M. Kriener, and S. Kittaka 
for useful discussions and for their assistance in measurements.
We also thank S. Maegawa, S. Fujimoto, S. Nakatsuji, and M. J. Lawler for fruitful discussions.
This work was supported by the Grant-in-Aid for the Global COE Program 
``The Next Generation of Physics, Spun from Universality and Emergence'' 
from the Ministry of Education, Culture, Sports, Science and Technology 
(MEXT) of Japan.
It has also been supported by Grants-in-Aid for Scientific
Research from MEXT and from the Japan Society for the
Promotion of Science (JSPS).

\bibliography{reference}
\end{document}